# The Bibliotelemetry of Information and Environment: an Evaluation of IoT-Powered Recommender Systems

Jim Hahn

*University Library, University of Illinois at Urbana-Champaign, USA. jimhahn@illinois.edu*

**ABSTRACT**

Internet of Things (IoT) infrastructure within the physical library environment is the basis for an integrative, hybrid approach to digital resource recommenders. The IoT infrastructure provides mobile, dynamic wayfinding support for items in the collection, which includes features for location-based recommendations. A modular evaluation and analysis herein clarified the nature of users' requests for recommendations based on their location and describes subject areas of the library for which users request recommendations. The modular mobile design allowed for deep exploration of users' bibliographic identifiers throughout the global module system, serving to provide context to the browsing data that are the focus of this study. Bibliotelemetry is introduced as an evaluation method for IoT middleware within library collections.

**KEYWORDS**

internet of things, IoT, bibliographic classification, academic libraries.

**INTRODUCTION**

The Internet of Things (IoT) is an advanced technology of novel connectivity that encompasses devices other than the classic desktop computing paradigm. IoT infuses internet affordances into objects, buildings, and other infrastructures. This technology is poised to affect library spaces, collection evaluation, and user services. In this research paper, IoT advances were the basis for an integrative, hybrid approach to digital resource access while navigating the physical library environment. According to David Rose, author of *Enchanted Objects* and researcher at the MIT Media Lab, "The Internet of Things is made possible through ubiquitous connectivity. An Internet connection allows the transmission of sensing and signalling information, the processing and storage of information, and the delivery of new services" (Rose 2014, p. 194-195). The authors of *Designing Connected Products* underscored the importance of IoT data stating, "The falling cost of sensing and computation means that there are more and more devices that are capable of capturing, and acting on, data about the world" (Rowland, Goodman, Charlier, Light, & Lui, 2015, p. 573). The scale of data sensor captures about collections and their use introduces a new research area for information science. A recent Library Technology Report, "The Internet of Things: Mobile technology and location services in libraries," detailed the use of a Bluetooth Low Energy (BLE) beacon array installed over an open undergraduate library collection. This infrastructure provides mobile, dynamic wayfinding support for items in the collection. It is dynamic because of its ability to show the users their location in real time as they move throughout the book stacks and the service includes features for location-based recommendations. The case reported on how the system used Library of Congress classification in nearby books to provide recommendations within the Minrva mobile app (Hahn, 2017a). Users were presented with recommendations to print items in popular circulation. The location of the mobile device was used to recommend such electronic items as databases, indices to journals, and e-books. A contention of location-based recommendation in library book stacks is that as users navigate the library's physical space, they navigate intellectual space through which topic discovery of both print and electronic items can be enhanced further (Hahn, 2011). Development and evaluation of IoT navigation services in print collections reference and extend foundations of information science concerned with recommending. In this research, we sought to enrich information science paradigms that reference classification for finding like items. We used API log analysis to evaluate IoT wayfinding and navigation support. This analysis will foster an understanding of the nature of users' requests for recommendations based on location. Because the recommendations provided were based on location, it is possible to ascertain within what section of the library users are requesting topic-based recommendations and to note the areas for which users were most likely to request recommendations.

**BACKGROUND**

IoT technology is poised to provide system designers, data scientists, and librarians with significant data from which to derive greater insight in collection analysis. IoT-powered services are beginning to be implemented in libraries, particularly in smart building development. Libraries have taken particular advantage of such mobile affordances as Bluetooth and Wi-Fi capabilities for IoT-powered wayfinding in library buildings. One recent exemplar is the University of Oklahoma's library, which began a NavApp project that allowed students to use mobile devices for indoor turn-by-turn guidance in the library and on campus. According to the University of Oklahoma Libraries NavApp website, "The OU Libraries' NavApp guides users throughout the Bizzell Memorial Library building, navigating them to resources, service desks, event spaces, and more, as well as guiding users to the libraries' branches and special collections" (University of Oklahoma, 2017). The project's indoor GPS functionality takes advantage of IoT affordances of Wi-Fi access points paired with mobile handheld technologies. The iBeacon library tours at Virginia Tech is a related project that uses IoT technology to explore information within a library building (Bradley et al., 2016). With Estimote iBeacons, the Virginia Tech Library could update self-guided tour applications to provide easier access to contextually relevant videos and information. The iBeacon tour was introduced to scale new student orientation. This is indicative of how IoT hardware will make it possible to provide services that scale to increasing work demands and provide contextually relevant information in libraries.

These contemporary projects, while technologically advanced, lead researchers to look for conceptual foundations. Have emerging technology projects in information science departed significantly from disciplinary foundations of information organization? Theoretical foundations of practice-oriented projects risk becoming superfluous because of their singular focus on technology qua foundations. However, technology is not an intellectual foundation for a discipline (Svenonius, 2000). Two overarching thematic frames are relevant to exploring the literature in this area: 1) conceptual frames of information behavior (Pettigrew, Fidel, & Bruce, 2001) and 2) frameworks related to browsing (Chang & Rice, 1993). Specific to browsing, a model for understanding was proposed to help clarify and expand its meaning, whereby it was argued that "… understanding browsing should include four major components—1) context, 2) influences, 3) browsing process, and 4) consequences – with iterations over time" (Chang & Rice 1993, p. 258). Additional frames synthesized in this research include those of Kuhlthau as well as Bates, whose works are situated in cognitive approaches to conceptual frameworks in information behavior. Other multifaceted approaches include "…multiple types of context such as the cognitive, social, and organizational context" (Pettigrew, Fidel, & Bruce 2001, p. 46).

### *Information Organization Foundations, Browsing, and Emerging Technology*

This work focuses on understanding browsing as a part of information behavior (Pettigrew, Fidel, & Bruce, 2001). Browsing is an aspect of information behavior that occurs under a variety of circumstances depending on the user context (McKay, Buchanan, & Chang, 2015). Both the berrypicking model of information search (Bates, 1989) and the information search process (Kuhlthau, 2004) are relevant and instructive models with which to begin to understand users' interactions as they search book stacks. For example, they may not necessarily want to go from point A to point B but rather follow an exploratory path through the book stacks. The chaotic nature of search was perhaps argued most compellingly by Dervin while theorizing about information design (Dervin, 2000). Dervin's work in this area—sense making—did not advance theories of sense making per se. However, the theory of information design is an illuminating model: Dervin proposed a conceptualization of information design as both order and chaos as a part of sense making. (Dervin, 2000).

Threads of order and disorder underscore a counterintuitive and underappreciated dichotomy in collections-based research: the collections themselves are ordered, but a user's research path is often uncertain (Eaton, 1991) and anything but orderly and standardized. In response to the classic article "Design of browsing and berrypicking techniques for the online search interface" the information science field began to appreciate and design information search interfaces to better address the nonlinear way users navigate information resources (Bates, 1989). Bates was influenced by researchers who also helped the information science field understand human-centered domains of search, rather than system-centric searching (Kuhlthau, 1988). Kuhlthau also provided guides to understanding user uncertainty while continuing to attend to the affective domains of discovery (Kuhlthau, 1993). And, in "Discovering information in context," Solomon developed a new research goal that proposed the need for information science research to encompass the variety of contexts in information search. These contexts look beyond interfaces and information systems, particularly with respect to how users make sense of information while navigating information environments (Solomon, 2002). Solomon's work on contextual information discovery underscores the practice-oriented approach of navigating library collections with IoT technology. The implications are relevant to IoT system design focused on supporting navigation. Solomon noted that "While many of the structures (e.g., categorizations, index languages, representations, metadata, displays) created by information professionals are designed to aid in access, discovery, and retrieval, they often fail in one or more of these functions" (Solomon, 2002, p. 240). Bibliotelemetry research was inspired by Solomon's call to support discovery in practice that upholds the non-linear, evolving nature of search as contextually significant.

Sustained research that focuses on user wayfinding in libraries is needed (Mandel, 2017). Mandel noted that future research questions for sustained study in this area include those that investigate how new technology can support wayfinding. Mobile technology-supported wayfinding in library practice must consider the evolving berrypicking approach that Bates introduced and upon which Solomon expanded. Lee's seminal work, which explored the nature of collections in the digital era, is instructive for system design of navigational IoT technology development. Lee argued compellingly for research on collections as information contexts (Lee, 2000). Our research considered both the tangible and intangible aspects of access and navigation of library collections with new IoT affordances in the digital age. This work was concerned with IoT's ability to provide enhanced, serendipitous discovery of library collections based on collocation (Svenonius, 2000). Empirical evidence helped verify that nearby print books on a shelf will circulate if neighboring books are checked out (McKay, Smith, & Chang 2014). Researchers also have used studies on book-stack browsing to inform digital e-book browsing systems (Hinze, McKay, Vanderschantz, Timpany, & Cunningham 2012). What this research addresses with IoT-enhanced wayfinding are the difficulties inherent to the tangibility of print collections. McKay et al. found that users have a range of needs while browsing: some need an exact item while others appreciate exploring a topic and finding items that may not be near their initial browsing location (McKay et al., 2015). Researchers have noted that "Systems could leverage topic classification schemes to offer readers a very few different-but-interesting books based on topic data: this approach would mimic the physical shelves, but, being data driven, would offer a higher chance of success" (McKay et al., 2015, p. 291). This study investigated the data affordances and limitations of subject-based filtering in library classification. A limitation of print layout is that while there are shelves of items arranged by subject, library collections are interconnected more deeply than physical shelves can indicate and consequently fail to achieve users' browsing-based navigation objectives. Subjects related from a user's perspective or needs are not always placed next to one another on shelves. Print arrangement includes electronic relations, wholly or in part, to resources not visible to the user navigating the stacks. IoT-powered services that take advantage of mobile devices paired with Bluetooth beacons and a set of interlocking middleware solutions can help address these inadequacies in the contextual issues of stacks navigation (Hahn, 2017a).

### *Modular Mobile Application Design with RESTful APIs*

The Wayfinder module within the Minrva mobile app was enhanced with a BLE beacon database to ascertain the location of the user device in the lower level of the Undergraduate Library at the University of Illinois. The Minrva app utilizes a modular design (Hahn & Ryckman, 2012). Each module is dependent on a unique bibliographic identifier and is not directly coupled with the larger system of modules. Modules communicate only indirectly. After a user searches for an item in the Catalog search module, they can then view a description of the item within the Display module, which will provide a summary along with location and availability information. Next, the user has the option to tap the Wayfinder module to be directed to the location of an item. Each module of the application uses a specific RESTful API. RESTful APIs are designed as concise, specifically formatted JSON data produced by servers to be consumed by other programs, such as mobile apps.

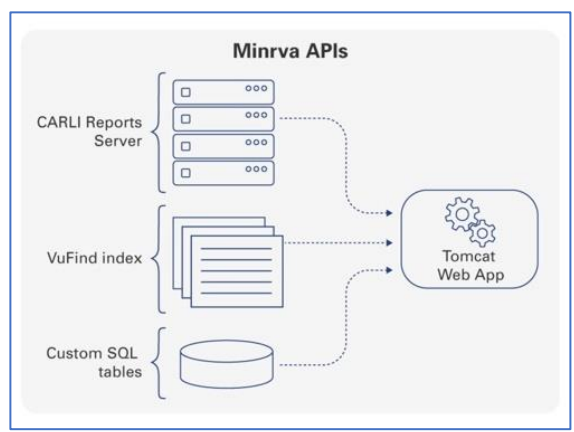

**Figure 1. The data sources the Wayfinder and location-based recommendation service require.**

The API developed for the Wayfinder module with recommendation support provides data based on three sources: basic call number layout (custom table we developed and curate), searching and filtering from the library catalog (VuFind), and ranking popularity (CARLI reports server). Examining Figure 1, we can see first the database labelled as the CARLI reports server refers to the Oracle Database that is the backend of the Integrated Library System (ILS). The system will request all data at once through one dynamic query. The goals and objectives for using Bluetooth beacons in the book stacks included the following: a) locating a user's device in the library book stacks, b) giving a user a recommendation for popularly circulating items, and c) showing the user relevant e-content based on their current location (Hahn, 2017a).

### *Recommendation Algorithm*

An algorithm, at its most basic and essential parts, is simply several functions that represent the steps the server is taking. Figure 2 details steps taken by the IoT middleware for gathering recommendation results defined by a pre-determined criterion that will be shown to the user based on the location of their device when they tapped the recommendation button (Hahn, 2017a). In order to process recommendations, the server will filter through functions shown in Figure 2 beginning with the current location. Based on patron location, we get the shelf ranges from our custom database of all call numbers. Once we know the ranges of a nearby stack we get relevant books and e-books. We filter for popularity based on that range and add any e-books that would be relevant. Finally, we add any online journal databases that may be relevant to the patron's starting location and serialize this as a JSON response. In order to gather database journal recommendations, the service relies upon an EBSCO article search API. While recommending is not the main purpose of the EBSCO API, the conceptualization of a "recommendation" is considered broadly to encompass those items more relevant by arrangement. The EBSCO API returns a set of results to a subject query (gathered from the call number). In the case that a single journal title holds most of the results, the service then recommends that journal as relevant to the subject area. The students still must search within the journal to obtain articles. This part of the algorithm, at its most basic, promotes the availability of topically relevant journal titles from the EBSCO API. Certainty there are limitations to this type of search since promoting some resources over others may leave out resources relevant but not found in the recommendation model (Hahn, 2017a).

```
// Get the patron's physical coordinates
PatronLocation location = getLocation(locationInfo);

// Get shelf ranges from Minrva DB, based off of patron location
ArrayList<String> ranges = getRanges(location);

// Get relevant books/ebooks based off of the shelf ranges
BooksDB bookdb = getEBooks(ranges, context);

// Filter and add popular books that are near the patron
List<RecBooksModel> books = bookdb.bookList;

// Filter and add relevant ebooks that would be shelved
// near the user, if the ebook was a physical item
List<RecBooksModel> bookEBook = getPopularBooks(ranges, books);

// Add relevant databases, based off of the patron's location,
// to the list
List<RecBooksModel> bookEBookDB = getDBSuggestions(bookdb.DBList, bookEBook);
```

**Figure 2. Demonstration of computing processes that researchers developed for the phone to receive the recommendations of print, e-books, and databases.**

**METHODS FOR BIBLIOTELEMETRY ANALYTICS**

Transactional search log analysis is a powerful tool that studies of the effectiveness of information systems employ (Jensen, 2006). Analysis of the server transaction logs recorded by library APIs is a comparable library research technique (Hahn, 2017b). Weblog analytics have proven to be a valuable method for understanding the information behavior of traditionally unobservable actions (Tobias & Blair, 2015). The focus of inquiry for log analytics in the case of this study was wayfinding, browsing, and selecting behavior in the library book stacks in general, and location-based recommendations, specifically. The analysis consists of extracting API logs, annotating the API logs, and then generating visualizations of the logs. Complete data are available from the University of Illinois databank (Hahn, 2018).

The Bibliotelemetry analysis is inspired by commonly used methods for analyzing transaction logs, where data are annotated or coded for analysis after logging transactions (Xie & Wolfram, 2009; Brett, German, & Young, 2015). This research focused on two sets of modular API logs from the server middleware between the users' mobile device and the Bluetooth beacon array in the undergraduate library book stacks. The middleware for the app was designed modularly such that the API that sends data to the phone for basic wayfinding is distinct from the API that sends recommendations based on a user location generated by the client device. Therefore, the two modular APIs were parsed, analyzed, and annotated as part of this research. IRB approval was obtained to evaluate mobile app use and user location stored in the API server logs. Splunk Enterprise software was used to parse server logs and Tableau was used to filter the data further and to visualize the results of subject distribution. An intermediary step from the Splunk Enterprise log file parsing was generating file annotations to plot subjects. An example of the filtered logs is shown below in Table 1. In Table 1, the Wayfinder API is denoted in the "uri" column and shows the targets of Wayfinding support, or basic collections navigation, without recommendations.

| uri | sum-records | X | Y | shelf-number | call-number |
|---|---|---|---|---|---|
| /api/wayfinder/map_data/uiu_undergrad/uiu_8127460 | 14 | 337 | 128 | 30 | PN1995 .C655 2015 |
| /api/wayfinder/map_data/uiu_undergrad/uiu_7419985 | 13 | 92 | 304 | 51 | SF446 .C763 2014 |

**Table 1. A sample from the annotated data generated from the middleware server/Wayfinder API.**

| uri | bib-id | bib-id | bib-id | bib-id | bib-id |
|---|---|---|---|---|---|
| /api/recommend/popularnear?x=4362.047852&y=3160.110596 | uiu_8378456 | uiu_7072382 | hat_817483 | uiu_7277188 | uiu_8375583 |
| /api/recommend/popularnear?x=3194.620605&y=4221.636719 | uiu_6548355 | uiu_7010231 | hat_819122 | uiu_6165435 | hat_1273444 |

**Table 2. A sample from the annotated data generated the middleware server/location-based recommendation API.**

The data from the server logs do not contain the subjects of items, but rather unique identifiers of bibliographic IDs. Researchers coded the files to associate the subject areas of the bibliographic IDs with an LC call number that would make possible a broader subject analysis of recommended items. An example of the subject/topic coding for call number association is shown in Table 3, which references and are directly sourced from Library of Congress Class lists (Library of Congress, 2018). The Library of Congress classification model is an imperfect model for organizing knowledge. However, a departure point for this research stems in part from the aphorism about the inherent fallibility of all models because when conceptual models are imperfect, it is more fruitful to scholarly inquiry to consider whether they are useful (Box, 1979). Consequently, while the LC classification model is not perfect, it is useful for IoT-powered recommendation evaluation and analysis. It has been applied uniformly in a standardized manner to print items in library collections.

| call-number | subject |
| --- | --- |
| B105.E9 G63 1974 | Philosophy (General) |

**Table 3. Recommended call number associated with subject.**

The study was conducted for one year, from September 1, 2016 through August 31, 2017. The collection for which the recommendations are provided included the print book stacks of a single floor of books in an undergraduate library. The collection of the undergraduate library numbers 125,000 monographs in open stacks. The goal of this evaluation was to understand the following: 1) Where are users when they request recommendations? 2) What are the subjects of the recommendations provided?

## RESULTS OF BIBLIOTELEMETRY ANALYTICS

During the course of the study, the Wayfinder module of the library mobile app recorded a total of 205 API hits to book item map support. There was a total of 431 API hits recorded in the server to the recommender middleware. To understand the use of location support, a heat map over a map of the undergraduate library was generated from the API logs used in this study, together with an open source python program that plots X, Y coordinates over a given image file.[1] The heat map indicates the popularity of a specific section: the darker the red, the higher the number of points reported at that location, while a lighter blue indicates less popularity. A reveal filter of lighter areas is displayed for the location of recommendation support.

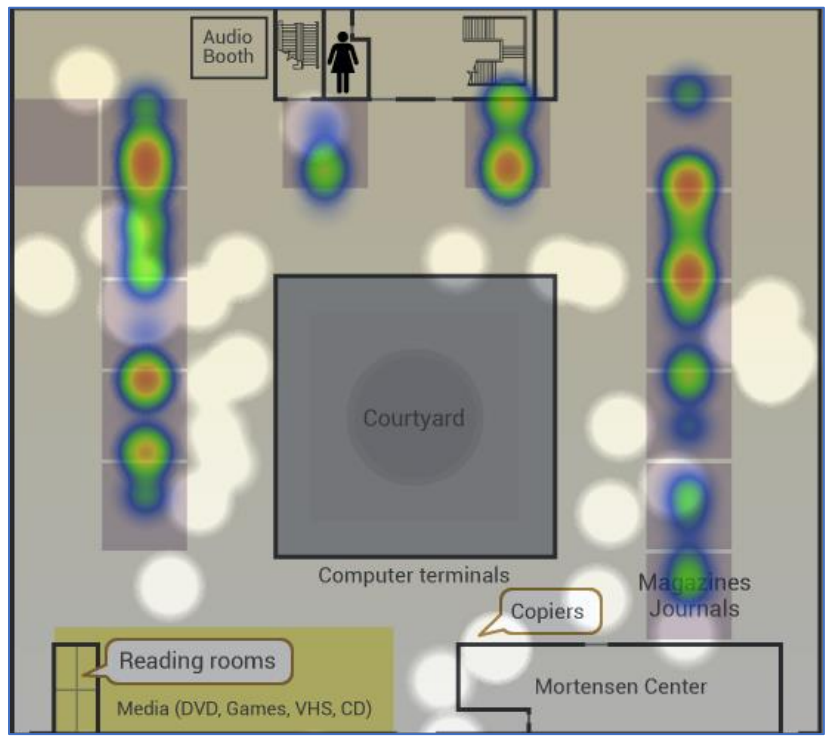

**Figure 3. Lighter areas are user location in the Undergraduate Library when requesting recommendations.**

To determine the topic areas of the wayfinding and recommendations, a subject distribution table was compiled with the API log data. The distribution is shown for the subjects of print items recommended based on user location (Figure 4). A long-tail power law model appeared to best fit the subject distribution findings (Anderson, 2006; Andriani & McKelvey, 2009). The results indicated that users of IoT-based recommendations are interested in a long-tail distribution of subjects. The secondary trend in literature topic areas may be interpreted as a combination of the following: 1) reflects the reading preferences and practices of those students, faculty, and staff researching literature; or shows the popularity of 2) leisure or recreational material overall in the collection; or 3) represents a larger distribution of subjects for this area compared with the collection overall. Users' antecedent Catalog searches, prior to viewing in the Display module were also a focus of data analysis since these were the steps taken by users prior to activating the Wayfinder module for locating an item and requesting recommendation support. Catalog and journal searches during the study are shown in Figures 7 and 8, respectively. Voyant tools (Sinclair & Rockwell, 2016) were used to generate text mining results. Results of the mining of words among both journal and catalog search modules are shown in Table 6.

---

[1] https://github.com/LumenResearch/heatmappy

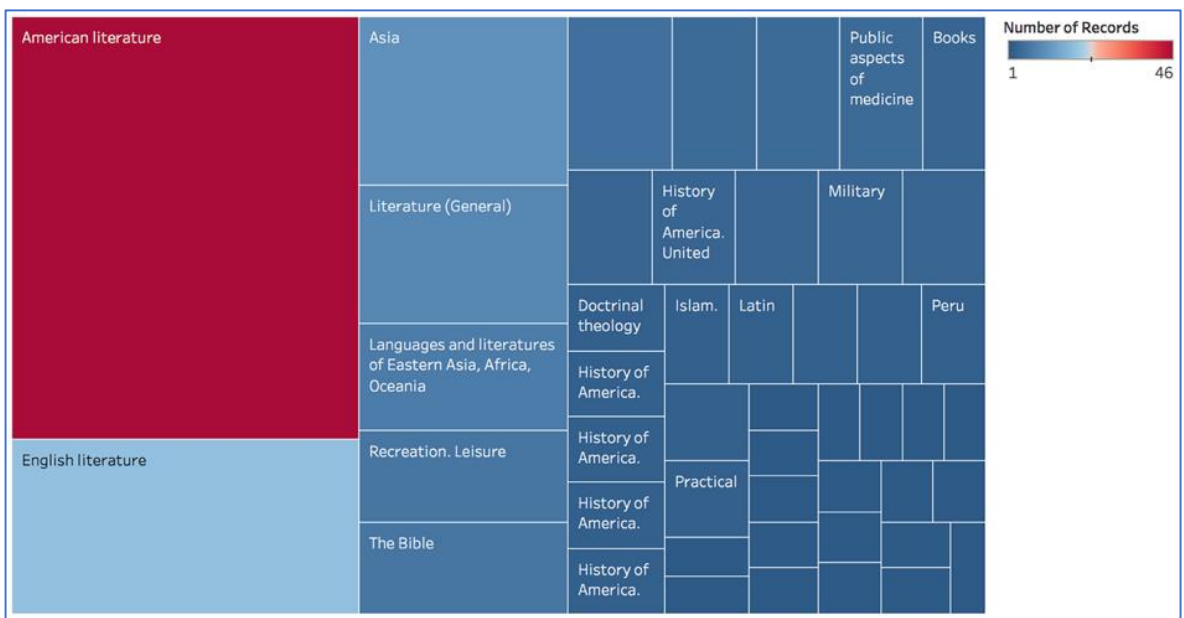

**Figure 4. Subjects of in-library items that were recommended by user location.**

*Bibliotelemetry: unique identifiers in the global module system*

The Bibliotelemetry analysis continues with the tracking user data of Wayfinder and BLE recommendations paired with the global modular mobile application data. This allows researchers the opportunity to follow a user from the Catalog Module search to the Display Module and finally to either Wayfinder and checkout or to Wayfinder and recommendations/checkout or more browsing afterwards. To begin to understand how satisfied users were with the recommendations and wayfinding support from a quantitative lens, researchers analyzed the bibliographic identifiers of circulation transactions after recommendation by the wayfinding module. Those items are detailed in Table 4. These are the print items recommended and subsequently checked out during the year under study. Recommendations for e-books in Table 4 do not have circulation records; however, the e-book access records are shown in Table 5. Further exploration of how bibliographic identifiers surface from catalog searching and into wayfinding throughout the module system is show in Figure 5 – which details how bibliographic identifiers from Wayfinder searching surfaced in other modules of the app, and Figure 6—which illustrates how identifiers from the location-based recommenders appeared in other modules throughout the app.

| bib-id | charge-date | call-number | subject |
|---|---|---|---|
| 7734446 | 8/31/17 9:49 | GV1469.62.D84 | Computer games. Video games. Fantasy games |
| 5180498 | 8/31/17 18:51 | PR9619.4.Z87 | English literature: Provincial, local, etc. |
| 4168504 | 8/31/17 18:52 | PZ7.R79835 | Fiction and juvenile belles lettres |
| 4060124 | 9/1/17 13:45 | PZ7 .R79835 | Fiction and juvenile belles lettres |
| 7878848 | 9/1/17 14:31 | PN1991.75 A24 | Radio broadcasts |

**Table 4. Bibliographic identifiers of checkouts after suggested to users within the Minrva Wayfinder module.**

| bib-id | req-date | call-number | subject |
|---|---|---|---|
| hat_606736 | 11/1/16 15:31 | PR85.C45 | English literature |
| hat_615138 | 11/1/16 15:32 | PS94.B5 | American literature |
| hat_345695 | 11/1/16 15:55 | E669.F37 | History of the Americas. United States |
| hat_659370 | 1/31/17 11:37 | E154.N38 | History of the Americas. United States |

**Table 5. Bibliographic identifiers of eBook views (e.g. clicks to open) within the location-based recommender.**

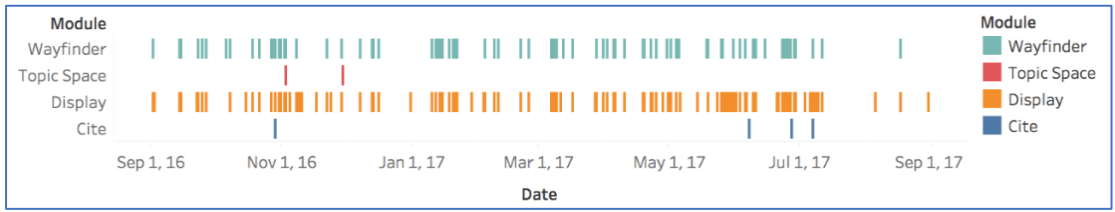

**Figure 5. Minrva modules that contained Bibliographic identifiers from Wayfinder logs.**

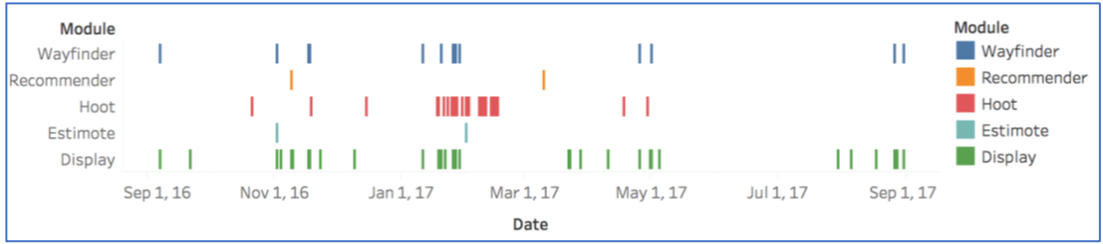

**Figure 6. Minrva modules that contained bibliographic identifiers promoted from location-based recommender.**

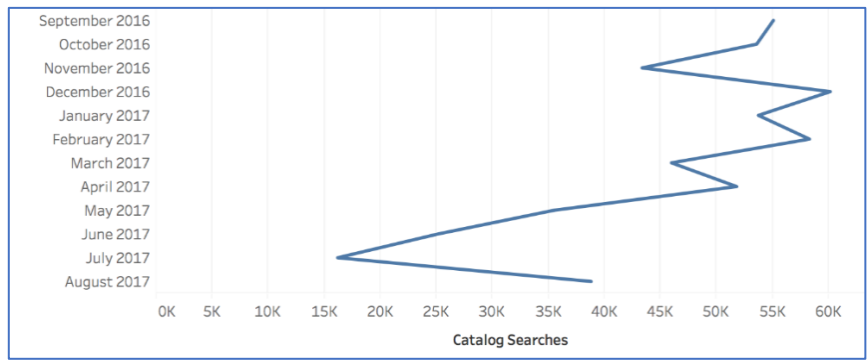

**Figure 7. Monthly catalog searches to the Minrva middleware.**

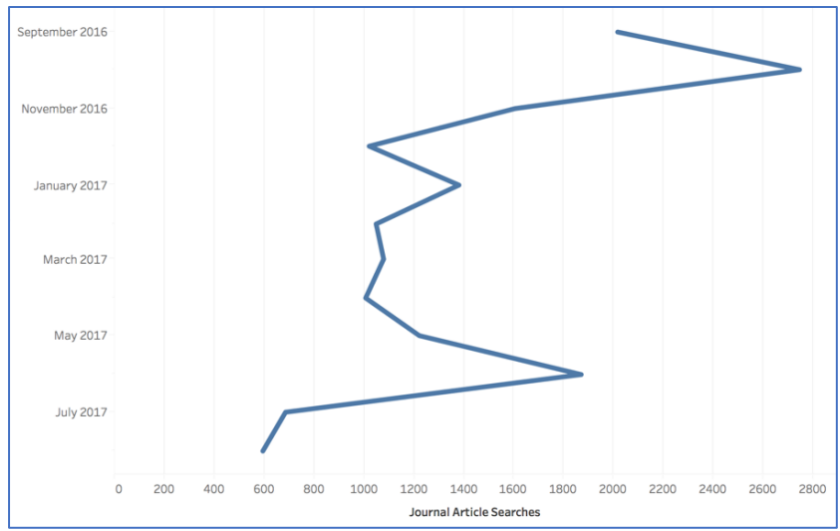

**Figure 8. Monthly journal article searches in the Minrva middleware.**

| Module | Total Words | Unique Word Forms | Most Frequent Words |
| --- | --- | --- | --- |
| Catalog Search | 1,951,982 | 3,016 | ps4 (41331); 2c (22982); trio (13514); daniel (12798); focus (12744) |
| Journal Search | 56,474 | 80 | fish (36232); google (4880); math (2016); test (1696); hiv (1692) |

**Table 6. Text mining of catalog and journal query word corpora.[2]**

---

[2] Note: Voyant Tools text mining (Sinclair & Rockwell, 2016) reports total words as a sum of every occurrence of every word, whereas the unique word forms are a summative word count based on first occurrence of a word only.

## FINDINGS

Initial checkouts and e-resource view findings (Tables 4 and 5) showed that users of the Wayfinder module seem to be interested in browsing more than navigating for checking out known items. Further Bibliotelemetry data within the global module system of the Minrva app showed that bibliographic identifiers of which users sought wayfinding support also surfaced within the Topic Space module and the Citation module. Citing an item may be an indication that the item was consequently used as a reference for a research paper. As per Figure 6, there were modules that contained bibliographic identifiers promoted from location-based recommender. Researchers could see that these identifiers showed up in users' course reserves interest (by way of the Hoot API logs—which alters users to the availability of in-library reserve texts) and appeared in the mobile app's account-based recommender, which suggests items to users based on their current checkouts. The significance of finding these identifiers within the global module system underscored the item's relative importance for use whether that included checkouts, e-access, or even, as an item that eventually became included in a professor's course reserves. While these findings may suggest a broader topic browse exploration by users, in the library stacks the qualitative nature of this exploration would be best explored through user interviews or focus groups as a companion study to complement this research thread.

An area for future qualitative study will focus on attending to the relative number of searches that were found (antecedent searches in catalog Figure 7 and temporally related journal article searches of Figure 8) versus the browsing observed by the Wayfinder module. The searching transactions far outnumbered browsing and checkout transactions in the study. In the future, researchers would specifically seek to understand why is it that searching is so prominent in the module system and to understand why browsing seemed important to users, but not necessarily for checkouts? What are the bookstacks offering users beyond their inventory? How do users decide to browse and not to select? Are these limitations in collection availability outside of the popular literature sections or is there another motivating set of factors? These may be best answered with focus groups and interviews. IoT navigational users' short-head preferences for literature may be used to inform the design of system improvements.

### *Long-Tail & Short-Head Findings*

In recommendations of print items, American Literature demonstrated the short-head trend as the subject area recommended most (Figure 4), with English Literature following in popularity. By analyzing and visualizing the IoT middleware server logs, it was possible to approximate where users were within the undergraduate collection when they requested recommendations (Figure 3). What is surprising to note about this figure is that most recommender system users did not actually appear to be in the stacks when they requested recommendations. The users were near book stacks when they requested suggested items but not within the boundaries of a book stack location, i.e. surrounded on either side by a shelf. The one exception was it appears that users requested recommendations within the boundaries of the literature and fiction areas of the book stacks. A future follow on study may continue to attend to the context of users entering or leaving the library. Because it appears that IoT services in this library may be the focus of leisure reading, designers may want to target their services to this subset of their collection. Specifically, these results may inform marketing and interface elements that would help users understand the features of the app that are amenable to traditional "reader's advisory" services. A further system design improvement can be seen in Figure 3: users appeared to request recommendations at the entrance and exit of the library. Since the system was designed to provide recommendations based on classifications associated with stack location, the library's entrances and exits may need to be reconsidered as they relate to stack-based recommendations.

While topic distribution in the short-head can show general trends in popularity, a long-tail distribution of subjects that occur less frequently deserves further interpretation, as the implications for service design and library marketing are not as straightforward. One group of researchers has shown that "Product categories with a higher number of titles and with a higher average demand display a shorter tail even with the same level of influence from the recommendation network. This is consistent with the conjecture that smaller categories with less popular products will have a more pronounced demand tail when influenced by recommendations" (Oestreicher-Singer & Sundararajan, 2012, p. 81). This observation may account for the short-head of subjects in fiction and literature. McKay found that day of the week effects browsing in the book stacks such that "Browsing is more likely on quiet days" (McKay, et al., 2015, p. 287). These time-sensitive data points consequently also effect the relevance of personalized recommendations shown to viewers of their service. Libraries that use a similar IoT-powered recommender may want to consider system design choices that use a combination of recommendations for short-head ranges and find ways to integrate less popular areas simultaneously based on profiling users through a collaborative filtering approach. Other library user data, such as further data mining subject data from circulation records, could be triangulated with the user findings here to be a weight in providing more personalized recommendations.

### *Limitations in Generalizing the Findings*

Observations that use system logs should be interpreted cautiously. Log analytic studies must consider two overarching limitations, such as interpretability and comparability (He, Qvarfordt, Halvey, & Golovchinsky, 2016). Researchers underscored the need to understand logs in context, using "… higher-level representation such as search tactics" (He et al. 2016, p. 1201). A review of recommender systems noted that, in some cases, content-based filtering "…causes overspecialized recommendations that only include very similar to those of which the user is already aware" (Park, Kim, Choi, & Kim, 2012, p. 10059). If the subject distribution is sparse, this may have been the case within literature when items circulating popularly may have been over-represented. Other digital library researchers have demonstrated ensemble approaches that combine a user's previous searches and views of items. System improvements also may be achieved by efforts to personalize the recommendations further with an individual's previous user movements, which would provide data for the system to learn and improve dynamically.

**CONCLUSION**

This paper detailed IoT methods for studying library spaces and collections. Trends in subject needs and users' preferences for topics can be identified by analyzing and visualizing IoT middleware server logs. One finding of this paper was the discovery of short-head occurrences of literature and fiction navigation in recommendation support. A long-tail finding showed significant diversity in the topics that IoT-powered location-based recommendation recommends to users in the library book stacks. The strength of this study was plotting the areas where users navigate in the stacks with IoT technologies, understanding where users are located when they are looking for assistance and, finally, providing a descriptive delineation of all topic areas of interest to the IoT-based wayfinding and navigation service users. Recommendation services in libraries can become more intelligent and powerful by taking advantage of streams of user motion data to provide recommendations. This study has helped fill descriptive gaps in the information science literature posed by Bates (1989) and others (Dervin, 2000; Solomon, 2002) to understand how users browse and navigate information environments.

Future research of long-tail subject recommendations is advisable to determine whether larger trends in the information science field may relate to other power laws. Explaining information behavior phenomenon using power law frames may help provide guidance in developing algorithms for recommendation and personalization services, which increase diversity of content, especially considering that filter bubbles have been shown to provide users more of what they already consume, while also serving to reinforce systemic bias (Noble, 2018). Therefore, a long tail of topic recommendations would be desirable from an information behavior perspective outside of recommender systems and within contemporary information environments generally.

**ACKNOWLEDGEMENTS**